\documentclass[aps,prl,reprint,superscriptaddress,showpacs]{revtex4-1}

\usepackage{graphicx}
\usepackage{epsfig} % use this package to include EPS format figures	
\usepackage{amsmath}
\usepackage{amsfonts}%
\usepackage{caption}
\usepackage[utf8]{inputenc}
\usepackage[english]{babel} %francais, polish, spanish, ...
\usepackage[T1]{fontenc}
\usepackage{color}
\usepackage[pdftex,colorlinks]{hyperref}
\usepackage{MnSymbol}

\begin{document}

\title{Experimental validation of a filament transport model in turbulent magnetized plasmas}
\author{D. Carralero}
\affiliation{Max-Planck-Institute for Plasma Physics, Boltzmannstr.~2,
 Garching, Germany}
\author{P. Manz}
\affiliation{Physik-Department E28, Technische Universit\"at M\"unchen,
James-Franck-Str.~1, Garching, Germany}
\author{L. Aho-Mantila}
\affiliation{VTT Technical Research Center of Finland, Helsinki, Finland}
\author{G. Birkenmeier}
\affiliation{Max-Planck-Institute for Plasma Physics, Boltzmannstr.~2,
 Garching, Germany}
\affiliation{Physik-Department E28, Technische Universit\"at M\"unchen,
James-Franck-Str.~1, Garching, Germany}
\author{M. Brix}
\affiliation{EUROfusion Consortium, JET, Culham Science Centre, Abingdon, OX14 3DB, UK.}
\author{M. Groth}
\affiliation{Aalto University, Espoo, Finland}
\author{H.W. Müller}
\affiliation{Max-Planck-Institute for Plasma Physics, Boltzmannstr.~2,
 Garching, Germany}
\affiliation{Institute of Materials Chemistry and Research, University of Vienna, Währingerstrasse 42, A-1090 Vienna, Austria}
\author{U. Stroth}
\affiliation{Max-Planck-Institute for Plasma Physics, Boltzmannstr.~2,
 Garching, Germany}
\affiliation{Physik-Department E28, Technische Universit\"at M\"unchen,
James-Franck-Str.~1, Garching, Germany}
\author{N. Vianello}
\affiliation{Consorzio RFX, C.so Stati Uniti 4,I-35127 Padova, Italy}
\affiliation{Ecole Polytechnique Fédérale de Lausanne (EPFL), Centre de Recherches en Physique des Plasmas (CRPP), CH-1015 Lausanne, Switzerland}
\author{E. Wolfrum}
\affiliation{Max-Planck-Institute for Plasma Physics, Boltzmannstr.~2,
 Garching, Germany}
\author{ASDEX Upgrade team}
\noaffiliation
\author{JET Contributors}
\affiliation{See the Appendix of F. Romanelli et al., Proceedings of the 25th IAEA Conference 2014, St Petersburg, Russia}
\date{\today}

\begin{abstract}
In a wide variety of natural and laboratory magnetized plasmas, filaments appear as a result of interchange instability. These convective structures substantially enhance transport in the direction perpendicular to the magnetic field. According to filament models, their propagation may follow different regimes depending on the parallel closure of charge conservation. This is of paramount importance in magnetic fusion plasmas, as high collisionality in the scrape-off layer may trigger a regime transition leading to strongly enhanced perpendicular particle fluxes. This work reports for the first time on an experimental verification of this process, linking enhanced transport with a regime transition as predicted by models. Based on these results, a novel scaling for global perpendicular particle transport in reactor relevant tokamaks such as ASDEX-Upgrade and JET is found, leading to important implications for next generation fusion devices.
\end{abstract}
%PACS: 52.35.Py (Macroinstabilities), 52.35.Ra(Plasma turbulence) 52.55.Fa (TOkamaks), 94.30.cq (magnetosphere MHD waves, plasma waves, and instabilities).
\maketitle	

Interchange instabilities are an ubiquitous phenomenon in magnetized plasmas where a density gradient coexists with a parallel force density of the same sign \cite{Chen65}. This instability leads to the formation of elongated structures known as filaments, flux tubes or prominences featuring a dipole electric polarization in the perpendicular plane. The resulting $E\times B$ drift propels the filament, greatly enhancing convective perpendicular transport. Examples can be found in disparate contexts, including astrophysical plasmas such as accretion discs \cite{Furukawa07} and planetary magnetospheres \cite{Southwood89}, where the destabilizing force is typically of the centrifugal type, and laboratory plasmas \cite{Endler95, Krash01}, where the force density typically comes from magnetic pressure gradients. This issue is particularly relevant for magnetically confined fusion plasmas, as it determines the propagation of filamentary structures, which have become recognized as the dominant radial transport mechanism in the region between the closed magnetic field lines and the wall, known as Scrape-off Layer (SOL) \cite{Boedo01, Krash07, Loarte07, Zweben07}. Therefore, filamentary transport strongly influences the parallel/perpendicular ratio of the particle and heat fluxes onto plasma facing components, thus determining the durability of plasma facing components and the sputtering of impurities from the main wall.\\

Basic models for filaments in fusion literature (see, e.g. Krasheninnikov \cite{Krash01}) describe how elongated structures propagate as the result of equilibration of plasma polarization caused by an effective gravity force: taking a reduced MHD approach on a magnetized stationary plasma under some arbitrary force density $\bf{F}$, the charge conservation equation can be expressed as
\begin{equation}
\nabla \cdot\frac{d}{dt}\biggl(\frac{nm_i}{B^2}\nabla_\bot \phi \biggr) =  \frac{1}{B} {\bf b}\cdot \nabla \times {\bf F} + \nabla_\parallel J_\parallel, \label{eq1}
\end{equation}
where $n$, $\phi$, $B$, $J$ and $m_i$ stand for density, potential, magnetic field, current density and ion mass, $d/dt = \partial/\partial t + ({\bf b} \nabla \phi)/B \cdot \nabla$ and ${\bf b} = {\bf B}/B$. Polarization (lhs of Eq. \ref{eq1}) is thus the result of an equilibrium between the drive (first term on rhs) and the parallel closure term (second term on rhs) accounting for the current parallel to $\bf{B}$. In the SOL context, ${\bf F}=2nm_ic_s^2/R \ {\bf e}_r$, stands for the effect of curvature and $\nabla B$ ($c_s$ is the sound speed and ${\bf e}_r$ indicates the radial direction). This can be easily generalized to many other forcing mechanisms such as the centrifugal force in Keplerian systems, gravity, etc. Equation \ref{eq1} yields different solutions depending on the parallel closure used for $J_\parallel$: the primary hypothesis in the SOL is that the filament extends along the field line to the solid wall, and $J_\parallel$ is thus limited by the sheath \cite{Krash07}. In this regime, known as ``Sheath Limited'' (SL), the polarization term is neglected and $\nabla_\parallel J_\parallel = \frac{2}{L_\parallel}nc_se(1-\exp{[-e(\phi-\phi_f)/T_e]})$ is canceled by the drive term \cite{Krash07}. Here, $\phi_f$, $e$, $T_e$ and  $L_\parallel$ are the floating potential, electron charge, electron temperature and parallel connection length. By solving Eq. \ref{eq1}, the perpendicular velocity of the filament, $v_b$, is then found to scale as
\begin{equation}
v_b=2c_s\frac{L_\parallel}{R}\biggl(\frac{\rho_s}{\delta_b}\biggr)^2 \propto \frac{1}{\delta_b^2}, \label{eq3}
\end{equation}
where $\rho_s$ and $R$ are the ion gyroradius and tokamak major radius, and $\delta_b$ is the filament perpendicular size. However, several phenomena, such as high SOL collisionality, large X-point magnetic shear, electromagnetic effects, etc. may invalidate the wall connection hypothesis, effectively disconnecting  the filament from the wall \cite{Krash07}. In this case, $\nabla_\parallel J_\parallel \rightarrow 0$ and polarization cancels the drive term, leading to a different regime, known as ``Inertial Regime'' (IN), in which \cite{Garcia06} 
\begin{equation}
v_b=c_s\sqrt{\tilde{p}_e\frac{\delta_b}{R}} \propto \delta_b^{1/2}, \label{eq4}
\end{equation}
where $\tilde{p}_e$ is the pressure in the filament normalized to the background value. These basic models have been refined by the inclusion of realistic SOL elements such as hot ions \cite{Manz13} and large fluctuation amplitudes (``full-f'' approximation) \cite{Wiesenberger14}, leading to improved versions of Eqs. \ref{eq3} and \ref{eq4}: 
\begin{equation}
v_{b}^{SL}= c_s (1+\tau_i)\frac{L_\parallel}{R}\frac{\tilde{n}}{\bar{n}+\tilde{n}}\biggl(\frac{\rho_s}{\delta_b}\biggr)^2; \ v_{b}^{IN}=c_s\sqrt{(1+\tau_i)\frac{\tilde{n}}{\bar{n}+\tilde{n}}\frac{\delta_b}{R}}, 
\label{eq5}
\end{equation}
where $\tau_i = T_i/T_e$, and $\bar{n}$ is the background density. Note that the isothermal limit has been assumed to approximate $\tilde{p} \simeq T \tilde{n}$. These scaling laws should be considered as upper boundaries, as any deviation from the pure interchange cross-phase between electric field and pressure fluctuations reduces the velocity of the blob with respect to predictions \cite{Nold12,Fuchert13}. Although these models have been successfully compared with experiment in basic plasmas \cite{Thelier09} and extensive characterization work has been made in tokamaks  \cite{Myra06b,Birk14,Manz15}, a direct measurement of the transition between the two regimes remains to be achieved in a fusion relevant plasma. Nevertheless, such transition has been invoked to explain the formation of the SOL density shoulder, i.e. a substantial increase of the far SOL density radial e-folding length, $\lambda_n\simeq (\nabla_r n/n)^{-1}$, observed in many tokamaks when a certain density is exceeded during L-mode operation \cite{Labombard01,Rudakov05,Garcia07}. Recent experiments in ASDEX Upgrade (AUG) \cite{Carralero14} linked the increase in $\lambda_n$ to an increase of the filament size and associated perpendicular particle transport in the outer midplane $\Gamma_{\bot,fil}$, which can be up to 40\% of the total transport after the transition according to comparison to EMC3-EIRENE simulations \cite{Tilman14}. This transition takes place as the collisionality increases in the SOL at the onset of divertor detachment, suggesting that the aforementioned disconnection might play a key role. Myra et al. \cite{Myra06} predicted this process in a two region model of the SOL using the effective collisionality $\Lambda$, 
\begin{equation}
\Lambda=\frac{L_\parallel/c_s}{1/\nu_{ei}}\frac{\Omega_i}{\Omega_e} \propto nT_e^{-3/2} \label{eq2}
\end{equation}
 as the control parameter, where $\nu_{ei}$ is the electron-ion collision rate and $\Omega_{i/e}$ stands for the gyrofrequency of ions/electrons. The disconnection takes place for $\Lambda>1$, when the characteristic parallel transport time is longer than the inverse of ion-electron collision frequency. Subsequent work indicates that this results in enhanced perpendicular transport as the consequence of increased filament velocities and creation rates \cite{Russell07}. Results from Ref. \cite{Carralero14} showed that the transition in AUG coincides with $\Lambda\simeq1$. The relation between filament regime transition and the shoulder formation has also been shown in JET \cite{Carralero14b}, where independent measurements of collisionality in the divertor and midplane regions suggested that only the collisionality in the divertor region could account for the disconnection process.\\

\begin{figure}
	%\centering
		\includegraphics[width=.9\linewidth]{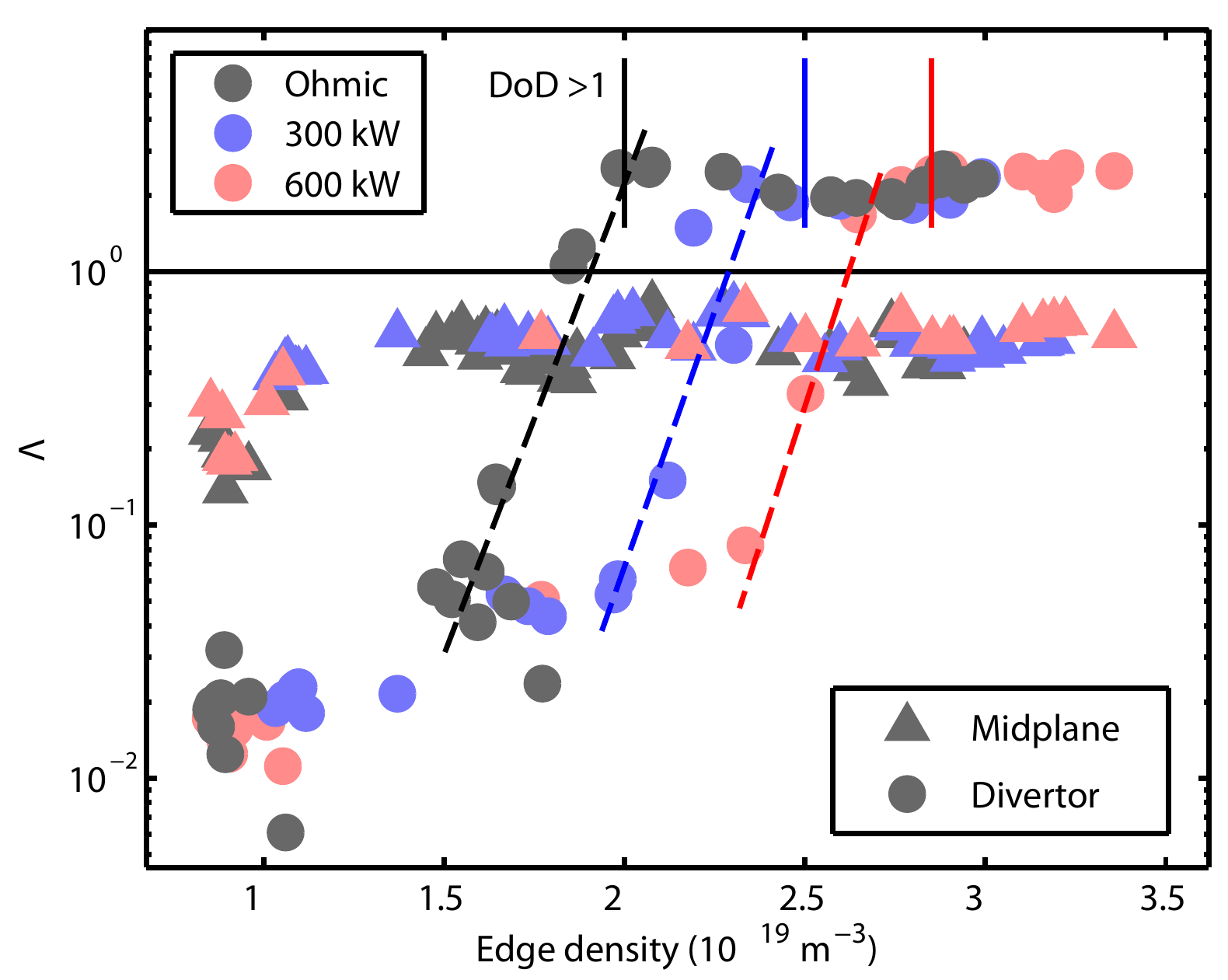}
	\caption{Collisionality parameter $\Lambda$ at midplane and divertor. Colors indicate different heating powers. The onset of divertor detachment is indicated by solid vertical lines.}
	\label{fig:1}
\end{figure}

The question of whether the collisionality in the divertor or in the midplane, $\Lambda_{div}$ or $\Lambda_{mid}$, are determining the shoulder formation is of great practical importance, as next generation tokamaks are foreseen to operate with partially detached, locally collisional divertors, while remaining hot and collisionless at the midplane. If shoulder formation is dominated by a filament transition induced by divertor collisionality, this phenomenon will not be reduced by a larger machine size, as it would according to other proposed mechanisms such as main wall recycling \cite{Lipschutz05}. In such case, a shoulder will probably be formed \cite{Carralero14b}, which would impact substantially main wall particle fluxes and erosion. To solve this conundrum, an experiment was designed in AUG to separate the contributions of $\Lambda_{mid}$ and $\Lambda_{div}$: in L-mode discharges, $T_e$ at the outer midplane far SOL does not depend strongly on the injected power $P_{IN}$, and remains nearly constant at $T_{e,mid} \simeq 20$ eV for a wide range of densities. In contrast, divertor detachment depends strongly on $P_{IN}$, occurring at lower midplane densities for lower heating power. Here, $\Lambda_{mid}$ and $\Lambda_{div}$ are defined as in Eq. \ref{eq2}, using respectively $n$ and $T_e$ Langmuir probe measurements from the midplane and target plates. In the case of $\Lambda_{mid}$, $L_\parallel$ is the connection length from the measurement point to the wall, $L_\parallel \simeq \pi R q_{95}$. In the case of $\Lambda_{div}$, it has been estimated as the connection length between the target and the X-point height at the flux surface of the measurements $L_\parallel \simeq \frac{1}{5} \pi R q_{95}$. This assumption is based on reciprocating probe and spectroscopic measurements of the cold region extension in front of the divertor at the time of the transition \cite{Carralero14, Potzel}. Therefore, by realizing a series of density ramps with different heating powers ($0$, $300$ and $600$ kW of ECH power, plus $500$ kW of Ohmic heating in all cases), $\Lambda_{mid}$ is kept constant while $\Lambda_{div}$ covers a range of more than two orders of magnitude, crossing the $\Lambda_{div} = 1$ critical point at different values of edge density, $\bar{n}_e$. This is shown in Fig \ref{fig:1}, where the collisionalities at the divertor and midplane for three $P_{IN}$ cases are displayed. While $\Lambda_{mid} \simeq 0.5$ for most of the $\bar{n}_e$ range, an exponential growth of $\Lambda_{div}$ can be observed at the onset of the divertor detachment (highlighted by colored dashed lines). The onset of detachment is marked by the DoD > 1 \cite{LoarteDoD}. Details on the diagnostic setup and data analysis are analogous to those described in Ref. \cite{Carralero14}.\\

\begin{figure}
	%\centering
		\includegraphics[width=1\linewidth]{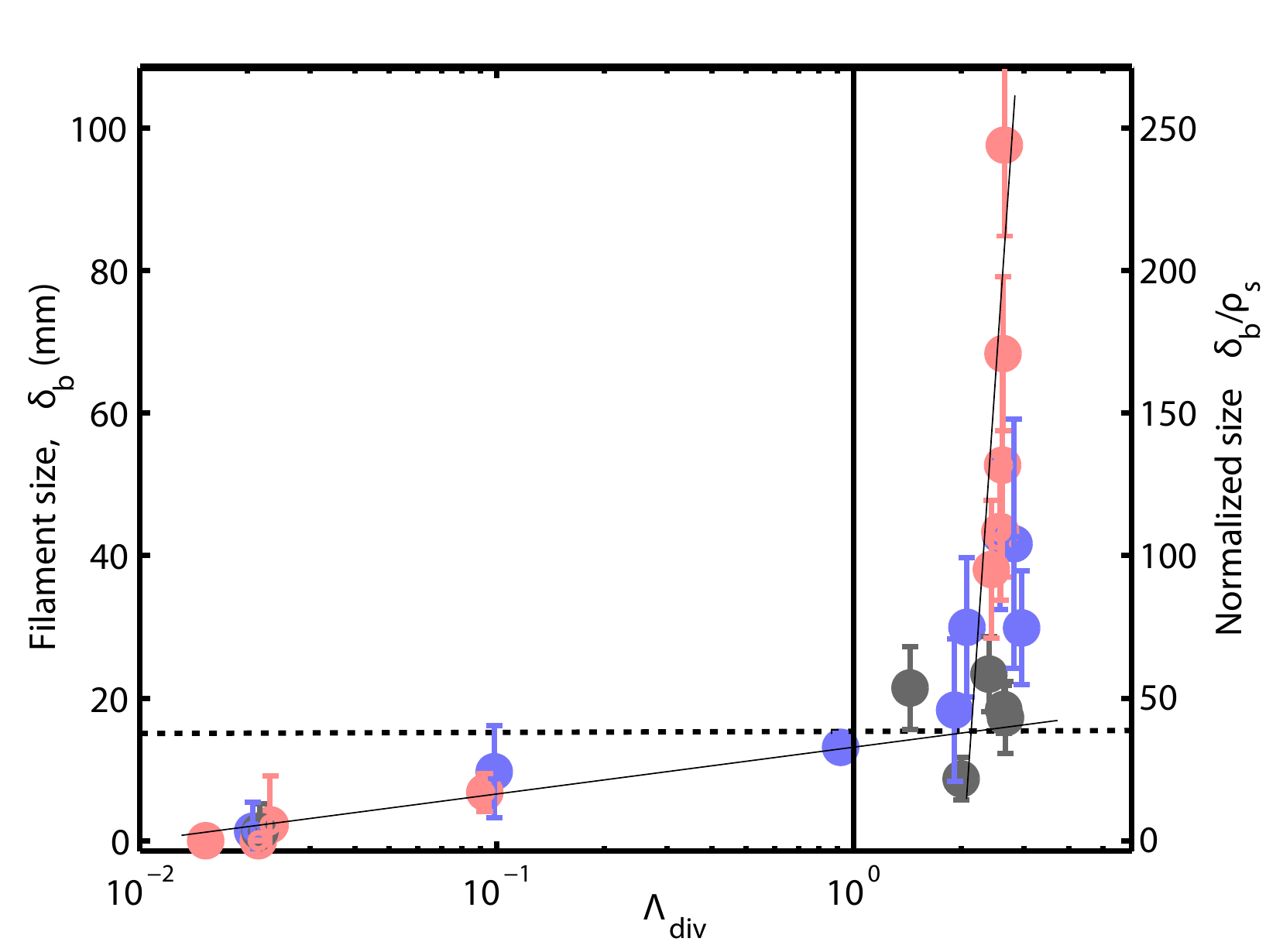}
	\caption{Size of filaments vs. collisionality. Colors indicate $P_{IN}$ as in Fig. \ref{fig:1}. Thin solid lines provided as guide to the eye to mark the two regimes.}
	\label{fig:2}
\end{figure}

\begin{figure}
	%\centering
		\includegraphics[width=1\linewidth]{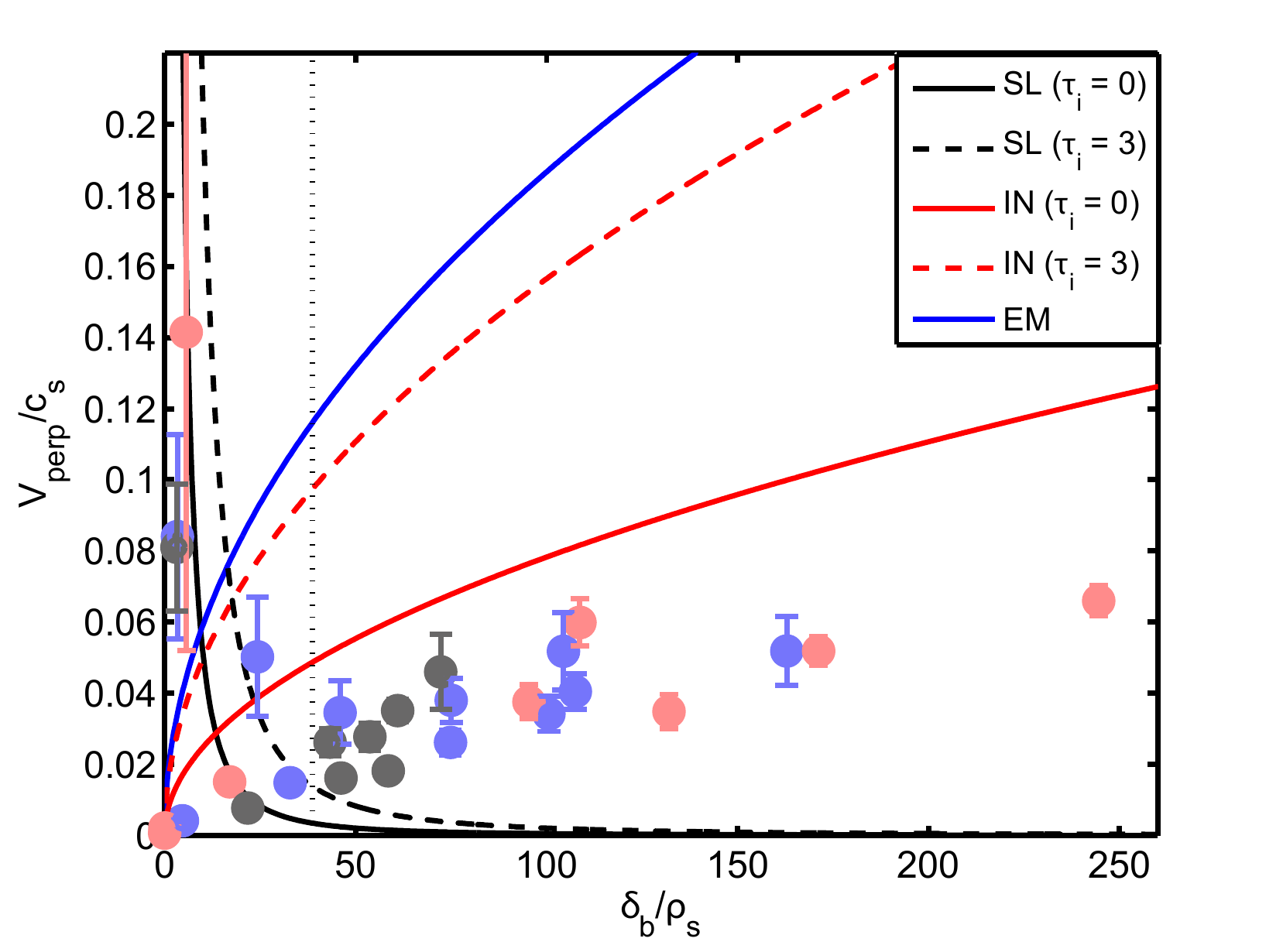}
	\caption{Filament scaling. Color code as in Fig. \ref{fig:1}. Black dotted line indicates the regime separation, as observed in Fig. \ref{fig:2}. Red/Black lines indicate IN/SL regimes, as expressed in Eq. \ref{eq5}. Solid/dashed indicates $\tau_i = 0$/$\tau_i= 3$. Blue solid line indicates the electromagnetic regime described in Ref. \cite{Manz13}.}
	\label{fig:3}
\end{figure}

Two different regimes appear approximately at each side of $\Lambda_{div} = 1$, as can be seen in Fig. \ref{fig:2}. For low collisionality, the size of filaments remains around one cm, and increases logaritmically with $\Lambda_{div}$. However, for dominant collisionality ($\Lambda_{div} > 1$), the size of filaments increases by up to an order of magnitude. The data set shows no dependence on $P_{IN}$, hence the transition is not determined by a $\bar{n}_e$ threshold, as can be deduced from the different $\Lambda_{div}(P_{IN},\bar{n}_e)$ paths seen in Fig. \ref{fig:1}. Furthermore, the velocity of small filaments follows the SL scaling $1/\delta_b^2$ and larger ones the IN scaling $\sqrt{\delta_b}$. This is shown in Fig. \ref{fig:3}, where the measured average filament velocity $v_\bot$ and size $\delta_b$ are normalized with $c_s$ and $\rho_s$ respectively as in Ref. \cite{Manz13}. Also, the upper bounds in the two regimes are represented by plotting the respective expressions of Eq. \ref{eq5}. In order to do this, the local magnetic field ($B_{SOL} = 1.4$ T), curvature radius ($R_{SOL} = 2.15$ m), $n_e$ and $T_e$ values at the probe position have been considered. To estimate $\tilde{n}$, the isothermal hypothesis has been extended to probe measurements, and the ion saturation fluctuation level $\tilde{j}_{i,sat}$ has been used as a proxy. Comparison of Fig. \ref{fig:3} with Fig. \ref{fig:2} reveals how both groups of data points correspond to those featuring $\Lambda_{div}<1$ and $\Lambda_{div}>1$: The horizontal dotted line crossing the elbow in Fig. \ref{fig:2}, is represented vertically in Fig. \ref{fig:3}, where it separates the two scalings.  Again, the same behaviour is seen for different $P_{IN}$ values, indicating the scaling is independent of this parameter (and thus also of $\bar{n}_e$). For both regimes, cold and hot ions have been considered ($\tau_i = 3$, as proposed in Ref. \cite{Manz13}). Filaments in the IN regime are better represented by the cold model. This is consistent with both theoretical analysis \cite{Manz15} and far SOL measurements carried out with field analyzers \cite{CarraleroPSI}, which suggest that filaments cool down after the transition. Data points in the lower left corner of Fig. \ref{fig:3}, corresponding to the smallest values of $\delta_b$, follow the IN scaling rather than the SL. This effect has been predicted \cite{Kube11} for SL filaments with $\delta_b \ll \delta_* \simeq 15\rho_s$, which would revert to an IN-like scaling.  Finally, the electromagnetic scaling proposed in Ref. \cite{Manz13}, in which electromagnetic effects dominate collisionality in the $\nabla_\parallel J_\parallel$ term, represents a worse fitting to the data than the SL and IN ones.\\

\begin{figure}
	%\centering
		\includegraphics[width=.9\linewidth]{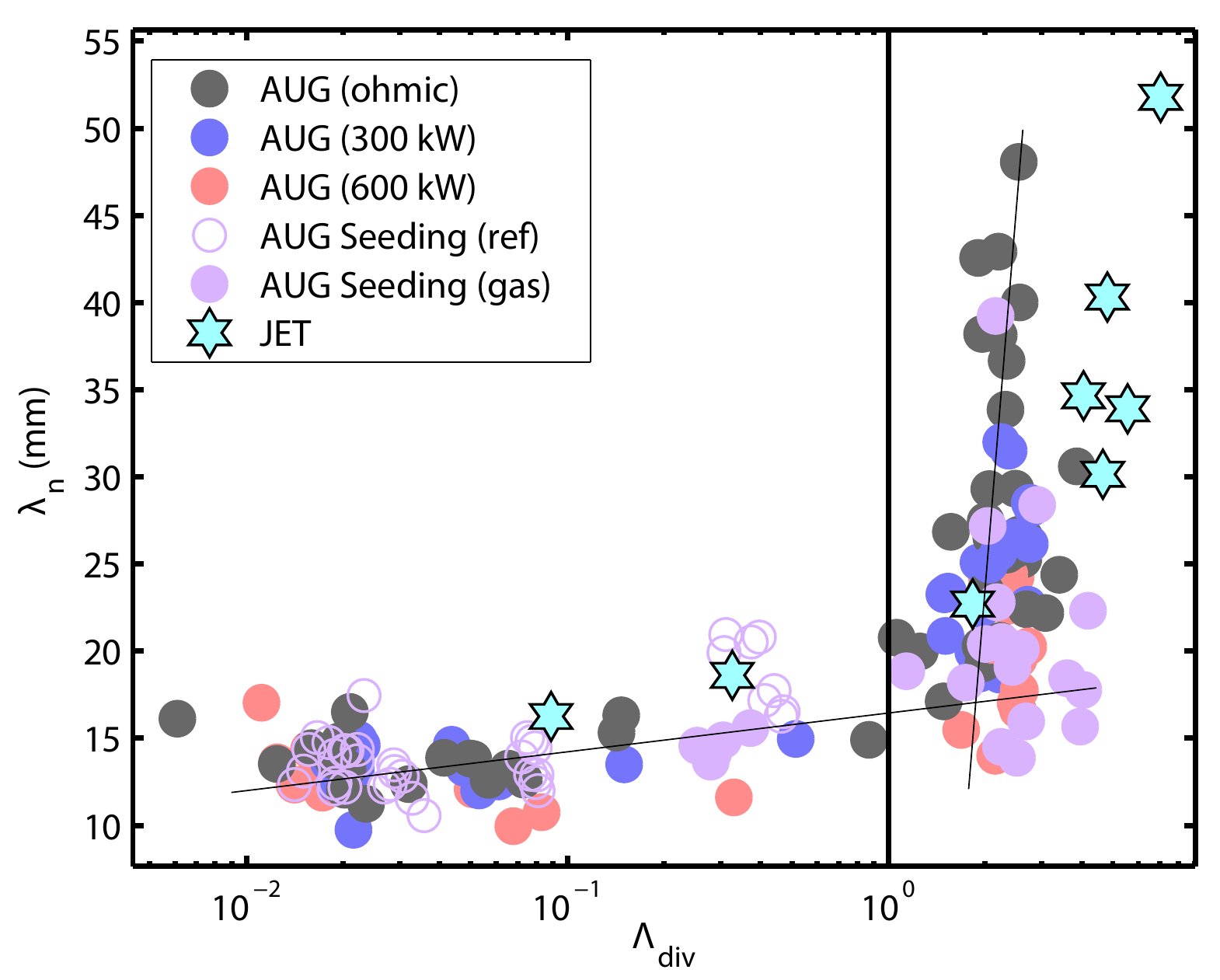}
	\caption{Density e-folding length scaling with divertor collisionality.  Black, blue and red circles represent $P_{IN}$ as in Fig. \ref{fig:1}. Purple circles correspond to seeded discharges (empty/solid mean before/after N$_2$ puffing). Light blue stars are data points from JET. Straight lines mark the two regimes as in Fig. \ref{fig:2}. }
	\label{fig:4}
\end{figure}

Finally, global perpendicular particle transport is found to be linked to divertor collisionality. In Fig. \ref{fig:4}, the density profiles remain almost constant until $\Lambda_{div} \simeq 1$ is surpassed. For $\Lambda_{div} > 1$, the SOL width rapidly increases by a factor of 3, indicating a substantial increase in $\Gamma_\bot$. As before, all parameter scans in AUG match onto one curve when $\Lambda_{div}$ is used as ordering parameter. Here, $\lambda_n$ is derived from Li-beam spectroscopy in the first $25$ mm outside the separatrix, as explained in Ref. \cite{Carralero14}. These results are independent on how detachment is achieved: If nitrogen is puffed in the divertor, local radiation cools down the region inducing detachment at a lower range of densities for a given $P_{IN}$. In Fig. \ref{fig:1}, seeding would cause a shift of the dashed line to the left compared to the unseeded discharge at the same $P_{IN}$. However, the $\lambda_n$ values from these discharges (empty/solid purple dots indicating before/after the puffing) scale exactly as the unseeded ones, despite the different $\bar{n}_e$ and $P_{IN}$ range (see Fig. \ref{fig:4}). Measurements in JET yield very similar results: $\lambda_n$ and $\Lambda_{div}$ are calculated for the discharges presented in \cite{Carralero14b}, using equivalent lithium beam and target probes data. The resulting data, displayed in Fig. \ref{fig:4} as light blue stars, display a remarkable similarity to the ones measured at AUG. As can be deduced from Fig. \ref{fig:1} this scaling is only obtained with $\Lambda_{div}$, and not with $\Lambda_{mid}$. This is consistent with previous studies showing that local collisionality at the midplane does not strongly influence midplane transport \cite{Lipschutz05}. Beyond magnetic fusion, these results can be regarded as the validation of a generic model for the propagation of structures resulting from interchange instabilities in magnetized plasmas. In this sense, the behavior of SOL filaments is strikingly similar to that of filamentary structures observed in the Jovian and Saturnian magnetospheres by the Galileo and Cassini spacecraft \cite{Rymer09, Thorne97}: The radial density profile of the equatorial iogenic plasma torus in the Jovian magnetosphere is dominated by the convection of ``isolated interchanging flux tubes'' driven by an isomorphical centrifugal force-driven interchange instability \cite{Hill81}. In this case, the parallel closure condition is determined by the conductivity at the end of the field lines, represented by the ionospheric conductivity at high planetary latitude instead of the plasma-wall interaction in the divertor region. As in the SOL case, when the coupling to the ionosphere is lost due to low conductivity, disconnected filamentary structures increase their radial speed, and, in qualitative agreement with Galileo measurements, the radial gradient is substantially decreased with respect to regions in which the equatorial plane and the poles are connected \cite{Bespalov06}.

Summarizing, the role of $\Lambda_{div}$ as a control parameter for the transition of both filament and global perpendicular transport regimes has been demonstrated, after separating its effect from other plasma parameters such as $\bar{n}_e$, $\Lambda_{mid}$ or $P_{IN}$. The velocity scaling of filaments changes from $1/\delta_b^2$ to $\sqrt{\delta_b}$ when the critical value $\Lambda_{div}=1$ is surpassed. These results represent strong evidence of filament propagation being governed by the mechanism presented in \cite{Myra06}, in which the interruption of the parallel circuit due to collisionality switches filaments from the SL to the IN regime. This change in SOL turbulence is accompanied by a substantial increase of the perpendicular particle flux $\Gamma_{\bot}$ in the far SOL, measured in two different tokamaks. The remarkable similarity of AUG and JET results supports the prediction of a shoulder formation in ITER. This has important implications for main wall particle fluxes and erosion in future fusion devices. Finally, the similarity between filaments driven by isomorphic instabilities in such disparate contexts as the SOL of fusion devices and planetary magnetospheres points towards a universal feature of transport in magnetized plasmas.\\

\textit{This work has been carried out within the framework of the EUROfusion Consortium and has received funding from the Euratom research and training programme 2014-2018 under grant agreement No 633053. The views and opinions expressed herein do not necessarily reflect those of the European Commission.}

\end{document}